\documentclass[iop,apj,appendixfloats]{emulateapj}
\usepackage{apjfonts}

\gdef\kms{km\,s$^{-1}$}
\gdef\msun{$M_{\odot}$}
\lefthead{van Dokkum et al.}
\righthead{}
\begin{document}

\title{Spectroscopic Confirmation of the Existence of Large, Diffuse
Galaxies in the Coma Cluster}

\author{Pieter G.\ van Dokkum\altaffilmark{1},
Aaron J.\ Romanowsky\altaffilmark{2,3},
Roberto Abraham\altaffilmark{4},
Jean P.\ Brodie\altaffilmark{3},
Charlie Conroy\altaffilmark{5},
Marla Geha\altaffilmark{1},
Allison Merritt\altaffilmark{1},
Alexa Villaume\altaffilmark{3},
Jielai Zhang\altaffilmark{4}}

\altaffiltext{1}
{Department of Astronomy, Yale University, New Haven, CT 06511, USA}
\altaffiltext{2}
{Department of Physics and Astronomy, San Jos\'e State University, San
Jose, CA 95192, USA}
\altaffiltext{3}
{University of California Observatories, 1156 High Street, Santa Cruz, CA 95064, USA}
\altaffiltext{4}
{Department of Astronomy \& Astrophysics, University of Toronto, 50 St.\ George St.,
Toronto, ON M5S 3H4 Canada}
\altaffiltext{5}
{Harvard-Smithsonian Center for Astrophysics, 60 Garden St., Cambridge, MA, USA}

\begin{abstract}

We recently identified a population of low surface brightness objects
in the field of the $z=0.023$ Coma cluster, using the
Dragonfly Telephoto Array.  Here we present Keck spectroscopy 
of one of the largest
of these  ``ultra-diffuse galaxies'' (UDGs), confirming that it is a
member of the cluster.
The galaxy has 
prominent absorption features, including the Ca\,{\sc{}II} H+K lines and
the G-band, and no detected emission lines.
Its radial velocity of $cz=6280 \pm 120$\,km/s is
within the $1\sigma$ velocity dispersion of the Coma cluster.
The galaxy has an effective radius of $4.3\pm{}0.3$\,kpc and a
Sersic index of $0.89\pm{}0.06$, as measured from Keck imaging. 
We find no indications of tidal tails or other distortions, at least out to
a radius of $\sim 2r_{\rm e}$.
We show that UDGs are located in a previously sparsely populated region of
the size -- magnitude plane of quiescent stellar systems, as they are
$\sim{}6$ magnitudes fainter than normal
early-type galaxies of the same size. It appears that 
the luminosity distribution of large quiescent galaxies is not continuous,
although this could largely be due to selection effects.
Dynamical measurements
are needed to determine whether the dark matter halos of UDGs are similar
to those of galaxies with the same luminosity or to those of galaxies
with the same size.

\end{abstract}


\section{Introduction}
In the Spring of 2014 we obtained wide-field ($3\arcdeg\times 3\arcdeg$)
observations of the Coma
cluster with the Dragonfly Telephoto Array ({Abraham} \& {van Dokkum} 2014). These images have
low spatial resolution ($\approx 6\arcsec$), but 
reach surface brightness
limits of $\mu(g)\sim 29.3$\,mag\,arcsec$^{-2}$. We found a population
of spatially-extended, low surface brightness objects in these images
({van Dokkum} {et~al.} 2015).
After combining the data with higher resolution imaging from the Sloan Digital
Sky Survey (SDSS) and the Canada France Hawaii Telescope (CFHT), we identified
47 objects with effective radii in the range $r_{\rm e}=3\arcsec - 10\arcsec$ and
central surface brightness $\mu(g,0)=24 - 26$\,mag\,arcsec$^{-2}$.

Based on the spatial distribution of the objects, and the smooth appearance of
one of them in a deep {\em Hubble Space Telescope} Advanced Camera for Surveys
image, we concluded that they are probably
galaxies in the Coma cluster. The Coma cluster has a radial velocity of
$cz = 7090$\,\kms\ ({Geller}, {Diaferio}, \& {Kurtz} 1999), and for a Hubble constant of 70\,\kms\,Mpc$^{-1}$
this implies a distance of $\approx 100$\,Mpc
($D_A=98$\,Mpc and $D_L = 103$\,Mpc). This distance places the galaxies in
an interesting region of parameter space: with effective radii
of $r_{\rm eff}=1.5$\,kpc -- $4.6$\,kpc
their sizes are similar to those of $\sim L_*$ galaxies, even though
their luminosities, colors, axis ratios, and {Sersic} (1968) indices are
similar to those of dwarf spheroidal galaxies. In {van Dokkum} {et~al.} (2015) [vD15] we
used the term ``ultra-diffuse galaxies'', or UDGs, to distinguish these
large, relatively round, diffuse objects from
the general classes of dwarf galaxies and low surface brightness galaxies.

Although it is plausible that most or all
of the 47 galaxies are indeed Coma cluster members,
this can only be confirmed with secure distance measurements. This is particularly
important for the largest galaxies: it may be that only the galaxies with the
smallest apparent sizes are at $\sim 100$\,Mpc, and that the largest ones are
considerably closer (see, e.g., {Merritt}, {van Dokkum}, \&  {Abraham} 2014).
Here we present spectroscopy and imaging
with the 10\,m W.M.\ Keck
Observatory of the largest galaxy in the vD15 sample, DF44.
The
goal of the spectroscopy is to test whether the galaxy is, in fact, in the
Coma cluster.  The imaging provides improved measurements of its
morphology and structural parameters.

\begin{figure*}[hbtp]
\begin{center}
\epsfxsize=16.5cm
\epsffile[0 0 1164 880]{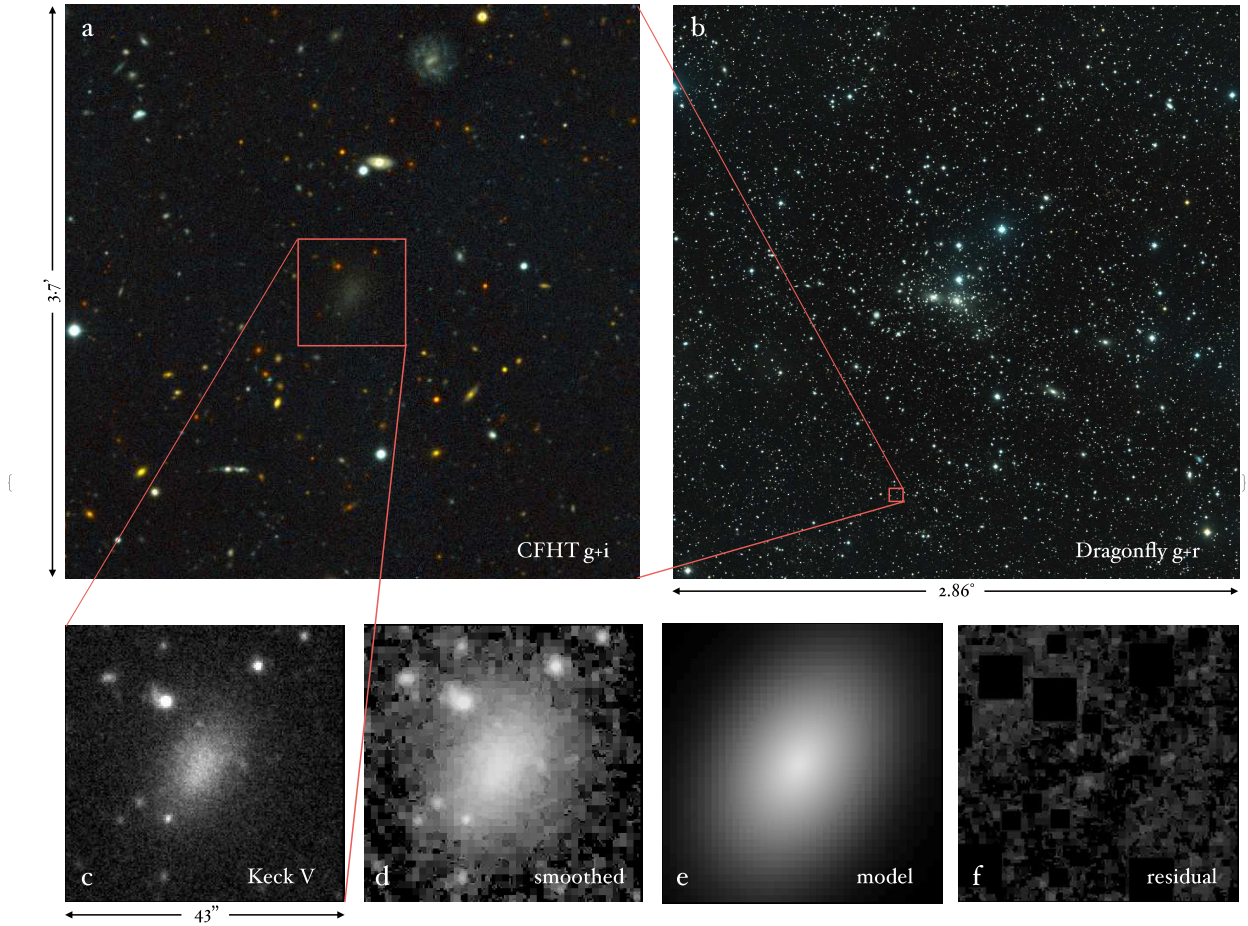}
\end{center}
\caption{\small
Imaging of DF44. Location of DF44 with respect to neighboring
galaxies (a; CFHT image) and with
respect to the center of the Coma cluster (b; Dragonfly discovery
imaging).
Panels c and d show the newly obtained Keck LRIS $V$-band
image, before and after adaptive smoothing.
Panels e and
f show the best-fitting single component
GALFIT model (Peng et al 2002) and the residual
after subtracting this model from d.
\label{im.fig}}
\end{figure*}

\section{Observations and Reduction}

Galaxy DF44 ($\alpha = 13^{\rm h}00^{\rm m}58.0^{\rm s}$;
$\delta = 26\arcdeg 58\arcmin 35\arcsec$)
was chosen because it is the largest (and second-brightest) galaxy
in the sample of vD15. It has 
a major-axis effective radius of $9\farcs 8$ and an integrated
apparent magnitude of $m_g=19.4$, as measured from CFHT images
(see Fig.\ \ref{im.fig}a).\footnote{DF44 is also featured in
Fig.\,1 of vD15. A careful reader may note that the location
of DF44 within the Coma field appears to be different in that paper.
The apparent location of DF44 in vD15
is, indeed, incorrect; the red square in
the Dragonfly image belongs
to another one of the four highlighted galaxies.}
The effective radius was derived by fitting a 2D Sersic
profile. The Sersic index was fixed to $n=1$ (exponential)
in this fit, as the S/N in the CFHT images is not sufficiently high to
measure $n$ reliably (see vD15). In Sect.\ \ref{structure.sec}
we update these measurements using deeper imaging from Keck.
As shown in Fig.\ \ref{im.fig}a,
DF44 does not have bright neighbors. It is
at a projected distance of $1\fdg 04$ from the center of the cluster, corresponding
to 1.8\,Mpc or $\sim 0.6 \times R_{200}$ ({Kubo} {et~al.} 2007).

The galaxy was observed 2014 December 18 and 19, using the Low Resolution
Imaging Spectrometer (LRIS; {Oke} {et~al.} 1995) on the Keck I telescope.
Conditions were mostly clear.
The 600\,lines\,mm$^{-1}$ grism blazed
at 4000\,\AA\ was used in the blue arm, and the gold-coated
1200\,lines\,mm$^{-1}$ grating blazed at 9000\,\AA\ in the red arm.
With a $1\farcs 5$ wide long slit, this configuration gives a
spectral resolution of $\sigma_{\rm instr}=2.5$\,\AA\ in the blue
and $\sigma_{\rm instr}=0.85$\,\AA\ in the red, corresponding to
$170$\,\kms\ at $\lambda=4500$\,\AA\ and $30$\,\kms\ at
$\lambda=8500$\,\AA. The slit was approximately aligned with the major
axis of the galaxy. 
The total integration time was 5400\,s, divided over
six exposures. The galaxy was moved along the slit in between exposures.

The blue and red spectra were reduced using standard techniques for
long slit spectroscopy (see, e.g., {van Dokkum} \& {Conroy} 2012, for an example using a similar
instrumental setup as employed here). Sky OH emission
lines were used for wavelength
calibration and rectification in the red. In the blue, an arc spectrum taken
immediately after the science exposures was used for this purpose.
The wavelength coverage was 3065\,\AA -- 5560\,\AA\ in the blue
and 7590\,\AA\ -- 9220\,\AA\ in the red. Sky subtraction was done
by fitting a linear function in the spatial direction. The galaxy was
masked in the fit.
One-dimensional spectra were
extracted by summing rows in the two-dimensional spectra. The extraction regions
are $\approx 11\arcsec$, and correspond approximately
to the rows where the flux is at least 30\,\% of
the peak. Extraction with
optimal weighting, or using a smaller or larger aperture,
does not improve the quality of the 1D spectrum.
A relative flux calibration was obtained
using observations of the
spectrophotometric standard Feige 110 ({Hamuy} {et~al.} 1992).

\begin{figure*}[htbp]
\begin{center}
\epsfxsize=16.5cm
\epsffile[40 177 772 655]{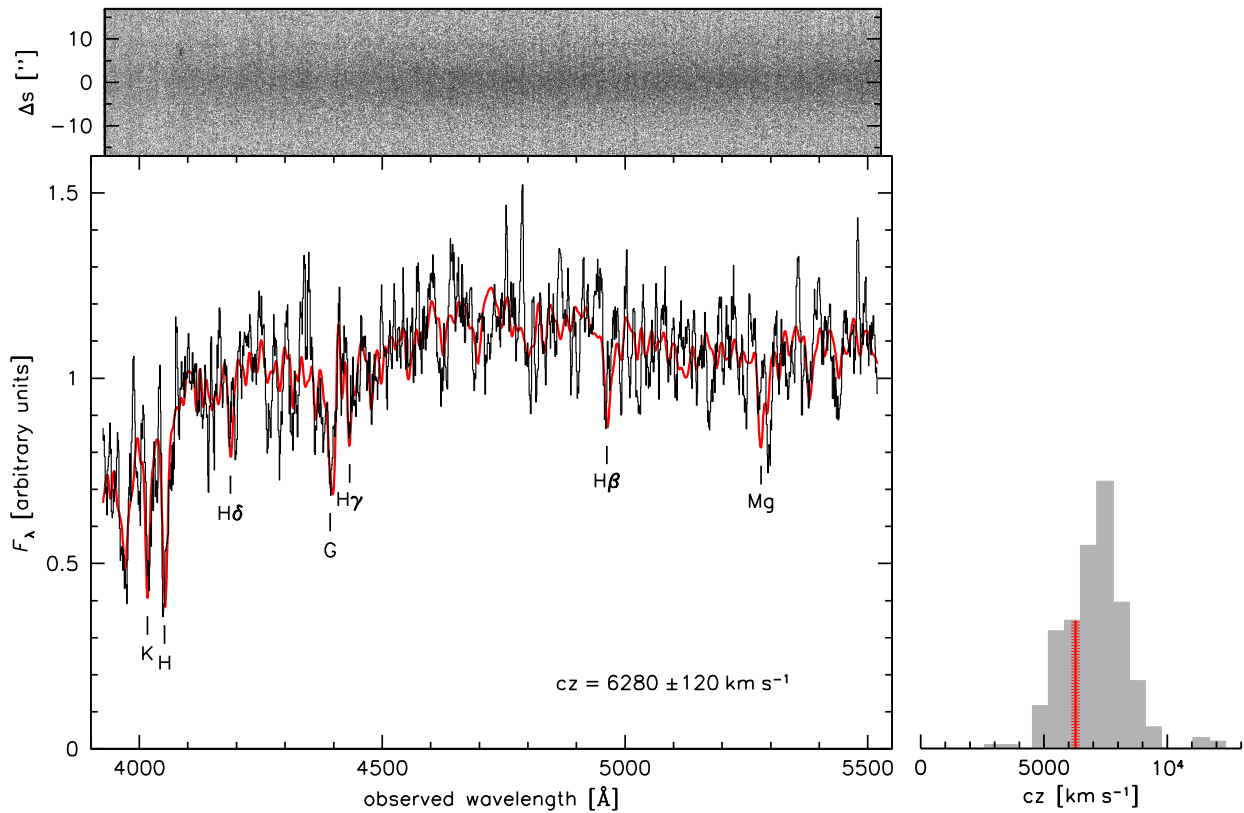}
\end{center}
\vspace{-0.3cm}
\caption{\small
Spectrum of DF44, obtained in 5400\,s with LRIS on the Keck I telescope.
Top panel: two-dimensional spectrum. The left axis
indicates the position along the slit, in arcseconds. Bottom panel: one-dimensional
spectrum, extracted from the 2D spectrum (black line). The spectrum was smoothed
with a 4.6\,\AA\ box car filter. The red line shows the smoothed
best-fitting model spectrum. The galaxy has an early-type
spectrum, and a redshift of $cz=6280 \pm 120$\,\kms.
The histogram on the right shows the redshift distribution of Coma, from
{Mobasher} {et~al.} (2001). The red line marks the redshift of DF44.
\label{spec.fig}}
\end{figure*}

We also obtained $V$ band imaging of DF44, using the blue arm of LRIS. A total
of 1080\,s was obtained over the two nights,
distributed over six dithered 180\,s exposures. The data were reduced using
standard techniques. In addition to a domeflat, a sky flat was used to
correct for remaining variation in the background. As the galaxy was
imaged on independent regions of the detector, the sky flat was
created from the six science exposures themselves. The FWHM seeing in the final,
combined image is $1\farcs 0$. The image was calibrated using SDSS
$g$ and $r$ photometry of stars in the DF44 field,  using
$V = g - 0.52(g-r) - 0.03$ ({Jester} {et~al.} 2005).

\section{Redshift measurement}

The LRIS spectrum of DF44 is shown in Fig.\ \ref{spec.fig}. Only the blue side
spectrum is shown, as the 
red spectrum has much lower signal-to-noise ratio (S/N) per resolution element.
No absolute calibration of the spectrum was
attempted, but the relative flux as a function of wavelength is
correct to $\sim 10$\,\% (as determined from the residuals between our
calibrated spectrum of Feige 110 and the one in Hamuy et al.\ 1992).
The spectrum resembles those of early-type galaxies: we
unambiguously identify the Ca\,{\sc ii} H+K lines, the G-band at 4300\,\AA, and
several other metal lines. The Balmer lines H$\beta$, H$\gamma$, and
H$\delta$ are also detected. No emission lines are found.

The redshift of DF44 was measured by cross-correlating the spectrum
with a range of templates of stars and galaxies, obtained from the
SDSS library.\footnote{http://www.sdss2.org/dr3/algorithms/spectemplates/index.html}
The best-fitting redshift is $cz = 6280 \pm 120$\,\kms.
The redshift distribution
in a $2.2$ degree$^2$ region of the Coma cluster  ({Mobasher} {et~al.} 2001)
is shown on the right.
The mean redshift of the
cluster is $cz=7090$\,\kms\ ({Geller} {et~al.} 1999), with a dispersion of $\sim 1100$\,\kms\
(e.g., {Colless} \& {Dunn} 1996; {Mobasher} {et~al.} 2001). We conclude that DF44 is a member of the Coma
cluster and is located at a distance of $\approx 100$\,Mpc.

We also fitted the spectrum with the flexible stellar population synthesis
(FSPS) models of {Conroy}, {Gunn}, \& {White} (2009), using the methodology of {Conroy} \& {van Dokkum} (2012).
This fit simultaneously determines the best-fitting
velocity dispersion, age, and metallicity,
along with the redshift. Unfortunately, the
S/N of the spectrum ($\approx 5$ per \AA)
is too low for stable constraints on these parameters, even when the
elemental abundance ratios are fixed to the Solar values. The red line
in Fig.\ \ref{spec.fig} is the best-fitting FSPS model, after matching
its continuum shape to that of DF44.

\begin{figure*}[hbtp]
\epsfxsize=14.0cm
\begin{center}
\epsffile{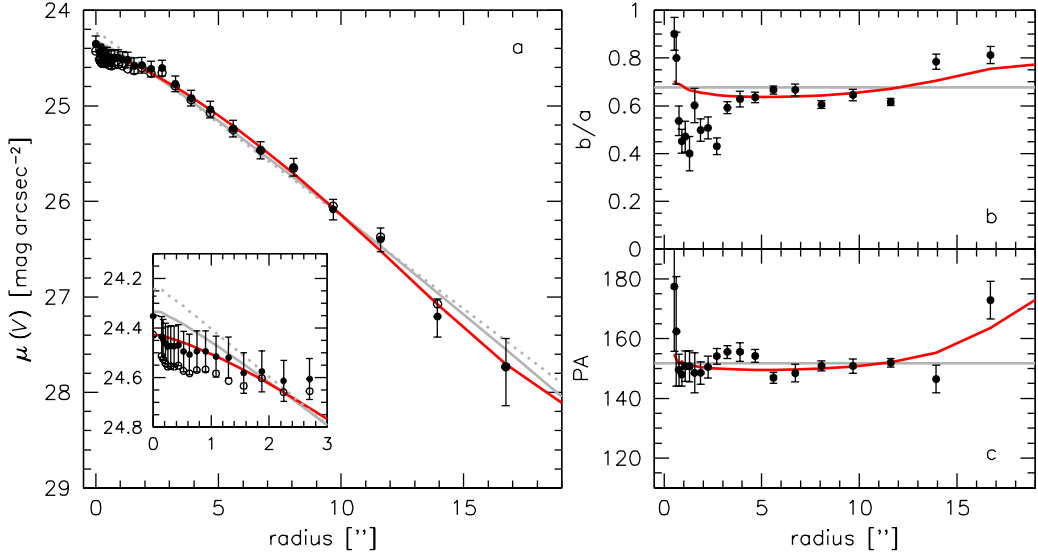}
\end{center}
\vspace{-0.4cm}
\caption{\small
Surface brightness (a), axis ratio (b), and position angle (c) of
DF44 as a function of distance along the major axis.
Open symbols are
the observed profile; solid symbols are corrected for the effects
of the PSF (see text). In all panels the
grey solid line is the best-fitting single component
2D Sersic model, with constant
axis ratio and position angle.
This model has a Sersic index $n=0.9$ and an effective radius $r_{\rm e} = 8\farcs 9$,
corresponding to 4.3\,kpc at the distance of DF44.
The grey broken line is an exponential fit, and the red line is
a two-component fit. The two-component fit has an effective radius of
$r_{\rm e}=8\farcs 4$.
The inset
shows the inner part of the surface brightness profile. In the central
$r<3\arcsec$ ($r<1.5$\,kpc) the
profile is depressed compared to the exponential fit.
\label{profile.fig}}
\end{figure*}

\section{Structure}
\label{structure.sec}

The Keck $V$ band image of DF44 is shown in Fig.\ \ref{im.fig}c. In
Fig.\ \ref{im.fig}d we show a version of the image that was
smoothed adaptively, to bring out the low surface brightness emission at
large radii. This smoothing was done for display purposes only; the
analysis was done on the original, unsmoothed image.
We note that the Dragonfly image 
of DF44 (see vD15) reaches fainter surface brightness levels than the Keck
image but is difficult to interpret
due to confusion with neighboring objects.
The galaxy has a regular, elliptical morphology, and
there is no evidence for tidal features,
spiral arms, or star forming regions, at least down
to $\mu(V)\sim 28$\,mag\,arcsec$^{-2}$.

We fit elliptical isophotes to
the image to measure the
surface brightness profile and to determine whether there is
evidence for isophotal twists or other irregularities.
Prior to the fit, all other objects in the image were masked carefully. The
sky background was determined from empty areas just outside of the region
displayed in Fig.\ \ref{im.fig}, and subtracted. The $1\sigma$
uncertainty in this background is approximately
$\mu(V)\sim 29$\,mag\,arcsec$^{-2}$ and is propagated into the errors in
the surface brightness profile.

The surface brightness profile of DF44 is shown in Fig.\ \ref{profile.fig}a.
The surface brightness is approximately constant at $\mu(V) \approx
24.6$\,mag\,arcsec$^{-2}$
within $r=3\arcsec$, and then falls off to reach $\mu(V) \approx
28$\,mag\,arcsec$^{-2}$ at
$r=20\arcsec$. The inner profile, highlighted in the inset,
is influenced by the point spread function (PSF). We corrected the profile
for the effects of the PSF following the procedure outlined in
\S\,3 of {Szomoru}, {Franx}, \& {van  Dokkum} (2012). First, a 2D {Sersic} (1968) model, convolved
with the PSF, was fitted to the image using GALFIT ({Peng} {et~al.} 2002). Then,
the residuals of this fit were added to an {\em unconvolved} 2D Sersic model,
and the surface brightness profile was measured from this PSF-corrected
image. The solid symbols in Fig.\ \ref{profile.fig}a show this PSF-corrected
profile.

The grey line shows the
best-fitting {Sersic} (1968) model.
Note that this model has a constant ellipticity and
position angle. It is a good fit to the
observed profile: the rms in the difference
between the solid points and the grey solid line
is 0.08 magnitudes.
The Sersic index of this model is $n=0.89 \pm 0.06$, and the best-fitting
effective radius is $r_{\rm e} = 8\farcs 9 \pm 0\farcs 6$. At the distance
of the Coma cluster of $D_A=98$\,Mpc this corresponds to $r_{\rm e} = 4.3\pm 0.3$\,kpc.
The total observed magnitude of the model is $m_V = 18.9$, and the absolute
magnitude is $M_V=-16.1$.
These results are consistent with our earlier measurement based on
shallower CFHT data; specifically, if
we force $n=1$ we find $r_{\rm e}=4.5$\,kpc, compared to $r_{\rm e}=4.6$\,kpc from
the CFHT data (vD15).

Panels b and c of Fig.\ \ref{profile.fig} show the variation in the
ellipticity and position angle as a function of radius. The radial
variation in  position is not shown, as 
the center of the best-fitting ellipse
is always within $1\arcsec$ of the mean position. There is some evidence
that the inner $r<3\arcsec$ ($r<1.5$\,kpc) is structurally distinct from
the rest of the galaxy: the surface brightness is depressed compared
to an exponential model (broken grey line; this depression
is why the best-fitting Sersic index is $0.9$ rather than 1)
and the galaxy appears more flattened
(axis ratio $b/a \approx 0.5$ versus $\approx 0.7$
at $r\gtrsim 5\arcsec$). The red line shows the result of
a two-component GALFIT fit, with the Sersic index of the
second component fixed to $n=1$.
The inner component has
a Sersic index of $n=0.69$, an effective radius of $r_{\rm e,i} =
7\farcs 5$, and an axis ratio $b/a=0.57$; the outer component
has $n\equiv 1$, $r_{\rm e,o} = 14\farcs 9$, and $b/a=0.73$.
The effective radius of the combined  model is $8\farcs 4$, very similar
to that of the single-component fit.

\section{Discussion}

The key result of this paper is the confirmation that one of the largest
UDGs in the field of Coma is a member of the cluster. In vD15
we had already argued that the 47 diffuse objects we discovered with the
Dragonfly Telephoto Array  are
very likely cluster members, but this was not based on direct distance
measurements. The objects with the largest apparent sizes are most
likely to be in the foreground,
and by confirming the distance to DF44 we can be
confident that most, and perhaps all, of the 47 galaxies are cluster members.

\begin{figure*}[hbtp]
\begin{center}
\epsfxsize=13.0cm
\epsffile[18 164 592 718]{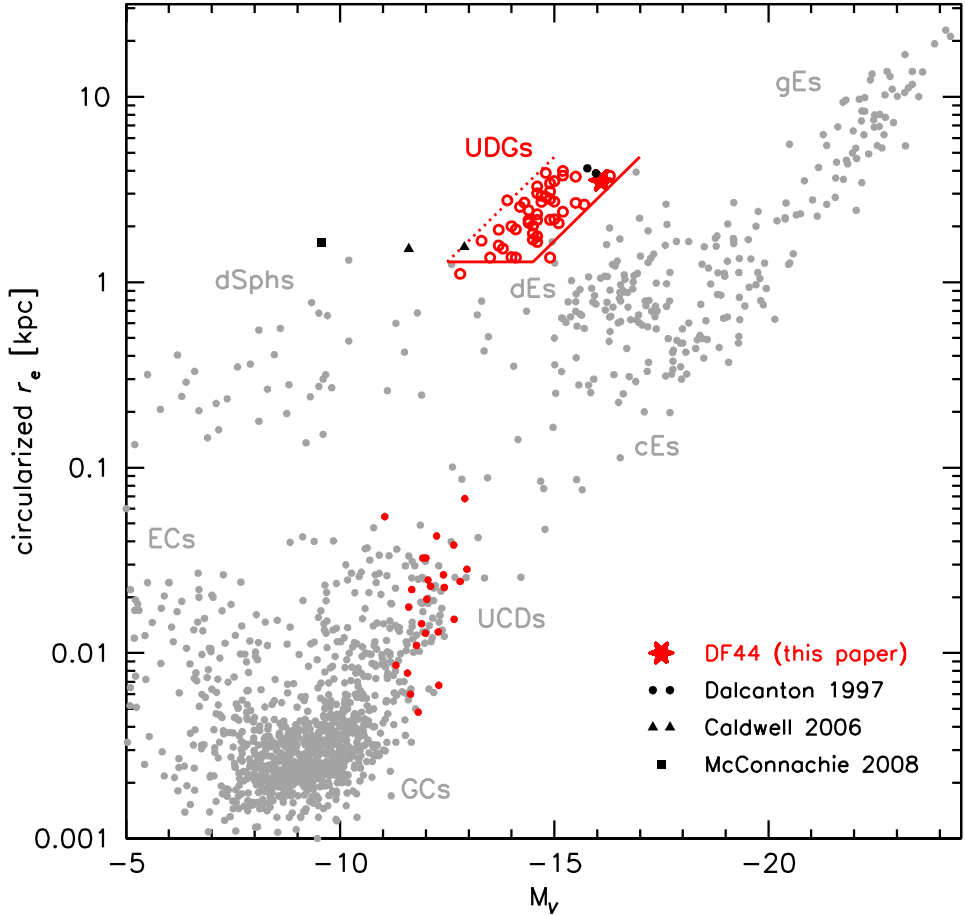}
\end{center}
\vspace{-0.3cm}
\caption{\small
Relation between projected circularized
half-light radius and absolute $V$ band magnitude for
quiescent
(early-type) objects, adapted from {Brodie} {et~al.} (2011). Solid symbols
denote distance-confirmed objects.
Red open circles
are ultra-diffuse galaxies from vD15,
assuming that they
are all members of the Coma cluster. 
The solid star is DF44. The red lines indicate the approximate
selection (solid) and detection (dotted) limits of vD15.
Red solid
circles are ultra-compact dwarfs 
in the Coma cluster ({Chiboucas} {et~al.} 2011), which have a factor of $\sim 10^7$ higher
3D stellar density
than the Coma UDGs. Black solid symbols are other diffuse galaxies
with confirmed distances
from the literature: the circles are the field galaxies
R-127-1 and M-161-1 ({Dalcanton} {et~al.} 1997), the
triangles are the Virgo cluster galaxies N\,lsb\,10 and SW2\,lsb31
({Caldwell} 2006), and the square is the Local Group galaxy
And XIX ({McConnachie} {et~al.} 2008). The grey point near DF44 is the Virgo
galaxy VCC 1661, which appears to be an erroneous measurement (McDonald
et al.\ 2011).
\label{context.fig}}
\end{figure*}

We note that this is not the first distance measurement to a large, diffuse galaxy.
{Caldwell} (2006) measured the distance to even fainter (but
also smaller) galaxies in the Virgo cluster from the location of the
tip of the red giant branch. {Dalcanton} {et~al.} (1997) measured redshifts of
seven large field galaxies with central surface brightness in the range
$\mu(V)=23-25$\,mag\,arcsec$^{-2}$. Two of these galaxies, R-127-1 and M-161-1,
may be similar to DF44: they
have no detected emission lines
and they have similar sizes and
surface brightness profiles.
And XIX is the only known example of a faint galaxy with $r_e>1.5$\,kpc in the
Local Group ({McConnachie} {et~al.} 2008).

In Fig.\ \ref{context.fig} we
place DF44 in context with these other UDGs, as well as with other classes
of quiescent (i.e., not star forming)
objects (see {Brodie} {et~al.} 2011).\footnote{Most of the data in this figure
come from the SAGES database (http://sages.ucolick.org/spectral\_database.html).}
The UDG data are taken from Table 1 in vD15, with $r_{\rm e, circ} = (b/a)^{0.5}
r_{\rm e, maj}$. With $b/a=0.68$, the
circularized effective radius of DF44 is 3.5\,kpc.
Solid red lines indicate the approximate UDG selection limits of
vD15, converted to
the axes of Fig.\ 4: $r_{\rm e, circ}>1.3$\,kpc and $\mu(V,0)>23.5$.
The broken red line indicates the approximate detection limit of
vD15 ($\mu(V,0)\lesssim 25.5$; this is driven by the depth of
the CFHT imaging that was used for confirmation).
UDGs highlight
the enormous range that exists in both axes of the size -- luminosity
diagram: at
their magnitude they
are a factor of $>100$ larger than ultra-compact dwarfs
and at their size they are a factor of $>100$ fainter than normal elliptical
galaxies. Interestingly all three classes of galaxies inhabit the same
environments and have broadly similar stellar populations.
A striking feature of Fig.\ \ref{context.fig} is the apparent gap
between the largest UDGs and giant elliptical galaxies of the same size
(that is, there are very few quiescent galaxies
with $r_{\rm e}> 3$\,kpc and $-20<M_V<-17$). We caution, however,
that the Brodie et al.\ sample was not designed to be complete
in this domain, and that such objects would fall outside
of the vD15 criteria.

The spectrum of DF44 does not provide new information on the formation
of UDGs, beyond confirming that they have an early-type spectrum
and no significant ongoing star formation. The deeper imaging enables
us to address one particular explanation for the existence of UDGs,
namely  that they appear large
because they have extensive tidal debris around them (see {Koch} {et~al.} 2012, for a
spectacular example of such a galaxy). Such tidal debris might be expected if
UDGs are in the process of being disrupted by the
tidal field of the cluster (e.g., {Moore} {et~al.} 1996; {Gnedin} 2003).
The Keck image of DF44 does not
provide evidence for this scenario: the 
galaxy does not appear to be in the process of
disruption, and the half-light radii are unlikely to be affected by
tidal features.
We note, however, that distortions may exist at fainter magnitudes.

The central depression in the surface brightness profile (relative
to an exponential profile) is a common feature in dwarf spheroidal
galaxies ({Irwin} \& {Hatzidimitriou} 1995; {McConnachie} 2012; {Merritt} {et~al.} 2014), and may be
evidence of the importance
of stellar feedback (e.g., {Read} \& {Gilmore} 2005; {Stinson} {et~al.} 2013).
It may be that this feedback suppressed
star formation at early times (see, e.g., {Oppenheimer} \& {Dav{\'e}} 2006; {Scannapieco} {et~al.} 2008; {Stinson} {et~al.} 2013), and that gas expelled in the
associated winds was swept up
in the  intra-cluster medium (ICM)
(e.g., {Abadi}, {Moore}, \& {Bower} 1999; {Mori} \& {Burkert} 2000) rather than falling back to the disk.
In this scenario Coma
UDGs could be considered ``failed'' $\sim L_*$ galaxies, that lost their gas to
the ICM.

Key to understanding UDGs is to know how much dark matter they have, and,
particularly, whether their halos resemble those of other galaxies of the
same size or those of other galaxies of the same luminosity.
The fact that UDGs are able to survive in the tidal field of Coma implies
that they are dark matter-dominated (see vD15), but
a quantitative mass measurement can only be obtained from internal kinematics.
We calculate the expected stellar velocity dispersion
as a function of radius using
simple spherical mass models with stars and {Navarro}, {Frenk}, \& {White} (1997) dark
matter halos, and a correlation between dark matter density and scale
radius as in {Spitler} {et~al.} (2012).
If the dark matter halos of UDGs are similar to those of dwarf galaxies
(with assumed $M_{\rm vir} = 6 \times 10^{10}$\,\msun)
their
luminosity-weighted
velocity dispersions are expected to be $\sim 35$\,\kms\ within the stellar
effective radius.
By contrast, if their
halos are similar to those of $L_*$ galaxies
(with assumed $M_{\rm vir} = 1.8 \times 10^{12}$\,\msun), their dispersions are $\sim
60$\,\kms.\footnote{In either scenario the {\em stellar} mass does not
contribute to the measured dispersion; the stars-only expectation for the
velocity dispersion is $\sim 7$\,\kms, for $M_{\rm stars}/L_V = 1.3$.}
In this paper we have shown what can be achieved in 1.5 hrs
with a traditional long-slit
spectrograph  on a large telescope. Using long exposure times with
low surface brightness-optimized integral field units (such as the planned
Keck Cosmic Web Imager; {Martin} {et~al.} 2010),
it should be possible to measure dynamical masses,
ages, and metallicities of these enigmatic objects.

\begin{acknowledgements}
We thank Nicola Pastorello for an independent check of the redshift
measurement.
Support from NSF grants AST-1312376, AST-1109878, and
AST-1211995 is gratefully acknowledged.
We also acknowledge the support of the Dunlap Institute, funded through
an endowment established by the David Dunlap family and the University
of Toronto.
\end{acknowledgements}



\begin{references}

\reference{} {Abadi}, M.~G., {Moore}, B., \& {Bower}, R.~G. 1999, \mnras, 308, 947

\reference{} {Abraham}, R.~G. \& {van Dokkum}, P.~G. 2014, \pasp, 126, 55

\reference{} {Brodie}, J.~P., {Romanowsky}, A.~J., {Strader}, J., \& {Forbes}, D.~A. 2011,  \aj, 142, 199

\reference{} {Caldwell}, N. 2006, \apj, 651, 822

\reference{} {Chiboucas}, K., {Tully}, R.~B., {Marzke}, R.~O., {Phillipps}, S., {Price}, J.,  {Peng}, E.~W., {Trentham}, N., {Carter}, D., {et al.} 2011, \apj, 737,  86

\reference{} {Colless}, M. \& {Dunn}, A.~M. 1996, \apj, 458, 435

\reference{} {Conroy}, C., {Gunn}, J.~E., \& {White}, M. 2009, \apj, 699, 486

\reference{} {Conroy}, C. \& {van Dokkum}, P. 2012, \apj, 747, 69

\reference{} {Dalcanton}, J.~J., {Spergel}, D.~N., {Gunn}, J.~E., {Schmidt}, M., \&  {Schneider}, D.~P. 1997, \aj, 114, 635


\reference{} {Geller}, M.~J., {Diaferio}, A., \& {Kurtz}, M.~J. 1999, \apjl, 517, L23

\reference{} {Gnedin}, O.~Y. 2003, \apj, 589, 752

\reference{} {Hamuy}, M., {Walker}, A.~R., {Suntzeff}, N.~B., {Gigoux}, P., {Heathcote},  S.~R., \& {Phillips}, M.~M. 1992, \pasp, 104, 533

\reference{} {Irwin}, M. \& {Hatzidimitriou}, D. 1995, \mnras, 277, 1354

\reference{} {Jester}, S., {Schneider}, D.~P., {Richards}, G.~T., {Green}, R.~F., {Schmidt},  M., {Hall}, P.~B., {Strauss}, M.~A., {Vanden Berk}, D.~E., {et al.} 2005, \aj, 130, 873

\reference{} {Koch}, A., {Burkert}, A., {Rich}, R.~M., {Collins}, M.~L.~M., {Black}, C.~S.,  {Hilker}, M., \& {Benson}, A.~J. 2012, \apjl, 755, L13

\reference{} {Kubo}, J.~M., {Stebbins}, A., {Annis}, J., {Dell'Antonio}, I.~P., {Lin}, H.,  {Khiabanian}, H., \& {Frieman}, J.~A. 2007, \apj, 671, 1466

\reference{} {Martin}, C., {Moore}, A., {Morrissey}, P., {Matuszewski}, M., {Rahman}, S.,  {Adkins}, S., \& {Epps}, H. 2010, in Society of Photo-Optical Instrumentation  Engineers (SPIE) Conference Series, Vol. 7735, Society of Photo-Optical  Instrumentation Engineers (SPIE) Conference Series, 0

\reference{} {McConnachie}, A.~W. 2012, \aj, 144, 4

\reference{} {McConnachie}, A.~W., {Huxor}, A., {Martin}, N.~F., {Irwin}, M.~J., {Chapman},  S.~C., {Fahlman}, G., {Ferguson}, A.~M.~N., {Ibata}, R.~A., {et al.} 2008, \apj, 688, 1009

\reference{} {McDonald}, M., {Courteau}, S., {Tully}, R.~B., \& {Roediger},
J. 2011, \mnras, 414, 2055

\reference{} {Merritt}, A., {van Dokkum}, P., \& {Abraham}, R. 2014, \apjl, 787, L37

\reference{} {Mobasher}, B., {Bridges}, T.~J., {Carter}, D., {Poggianti}, B.~M., {Komiyama},  Y., {Kashikawa}, N., {Doi}, M., {Iye}, M., {et al.} 2001, \apjs, 137, 279

\reference{} {Moore}, B., {Katz}, N., {Lake}, G., {Dressler}, A., \& {Oemler}, A. 1996,  \nat, 379, 613

\reference{} {Mori}, M. \& {Burkert}, A. 2000, \apj, 538, 559

\reference{} {Navarro}, J.~F., {Frenk}, C.~S., \& {White}, S.~D.~M. 1997, \apj, 490, 493

\reference{} {Oke}, J.~B., {Cohen}, J.~G., {Carr}, M., {Cromer}, J., {Dingizian}, A.,  {Harris}, F.~H., {Labrecque}, S., {Lucinio}, R., {et al.} 1995, \pasp, 107, 375

\reference{} {Oppenheimer}, B.~D. \& {Dav{\'e}}, R. 2006, \mnras, 373, 1265

\reference{} {Peng}, C.~Y., {Ho}, L.~C., {Impey}, C.~D., \& {Rix}, H.-W. 2002, \aj, 124, 266

\reference{} {Read}, J.~I. \& {Gilmore}, G. 2005, \mnras, 356, 107

\reference{} {Scannapieco}, C., {Tissera}, P.~B., {White}, S.~D.~M., \& {Springel}, V. 2008,  \mnras, 389, 1137

\reference{} {Sersic}, J.~L. 1968, {Atlas de galaxias australes} (Cordoba, Argentina:  Observatorio Astronomico, 1968)

\reference{} {Spitler}, L.~R., {Romanowsky}, A.~J., {Diemand}, J., {Strader}, J., {Forbes},  D.~A., {Moore}, B., \& {Brodie}, J.~P. 2012, \mnras, 423, 2177

\reference{} {Stinson}, G.~S., {Brook}, C., {Macci{\`o}}, A.~V., {Wadsley}, J., {Quinn},  T.~R., \& {Couchman}, H.~M.~P. 2013, \mnras, 428, 129

\reference{} {Szomoru}, D., {Franx}, M., \& {van Dokkum}, P.~G. 2012, \apj, 749, 121

\reference{} {van Dokkum}, P.~G., {Abraham}, R., {Merritt}, A., {Zhang}, J., {Geha}, M., \&  {Conroy}, C. 2015, \apjl, 798, L45

\reference{} {van Dokkum}, P.~G. \& {Conroy}, C. 2012, \apj, 760, 70

\end{references}
\end{document}